\documentclass[
reprint,
superscriptaddress,
showkeys,
 amsmath,amssymb,
 aps,
pre,
byrevtex
]{revtex4-2}

\usepackage{graphicx}% Include figure files
\usepackage{dcolumn}% Align table columns on decimal point
\usepackage{bm}% bold math
\usepackage{hyperref}
\hypersetup{
  bookmarksopen = true, 
  bookmarksnumbered = true, 
  colorlinks   = true, %Colours links instead of ugly boxes
  urlcolor     = blue, %blue Colour for external hyperlinks
  linkcolor    = red, %[RGB]{230 0 0}, %Colour of internal links.
  citecolor    = blue %Colour of citations
}
\usepackage{siunitx}
\DeclareSIUnit\molar{\mole\per\cubic\deci\metre}
\DeclareSIUnit\Molar{\textsc{m}}

\renewcommand{\arraystretch}{1.1}

\newcommand*{\VC}[1]{\bm{\mathrm{#1}}}

\DeclareMathOperator{\Tr}{Tr}

\begin{document}

\preprint{APS/123-QED}

\title{Ligand-induced incompatible curvatures control ultrathin nanoplatelet polymorphism 
and chirality}% Force line breaks with \\

\author{Debora Monego}
    \thanks{These authors contributed equally to this work.}
    \affiliation{ARC Centre of Excellence in Exciton Science, School of Chemistry, University of Sydney, Sydney, New South Wales 2006, Australia}
    \affiliation{The University of Sydney Nano Institute, University of Sydney, Sydney,
New South Wales 2006, Australia}

\author{Sarit  Dutta}
    \thanks{These authors contributed equally to this work.}
    \affiliation{Univ. Lyon, ENS de Lyon, CNRS, Laboratoire de Chimie, F-69342, Lyon, France}
    
\author{Doron Grossman}
    \affiliation{LadHyX, CNRS, Ecole polytechnique, Institut Polytechnique de Paris, 91128 Palaiseau Cedex, France}

\author{Marion Krapez}
    \affiliation{ARC Centre of Excellence in Exciton Science, School of Chemistry, University of Sydney, Sydney, New South Wales 2006, Australia}
    \affiliation{The University of Sydney Nano Institute, University of Sydney, Sydney,
New South Wales 2006, Australia}

\author{Pierre Bauer}
    \affiliation{Univ Lyon, Université Claude Bernard Lyon 1, CNRS, Institut Lumière Matière, F-69622, Villeurbanne, France}

\author{Austin Hubley}
    \affiliation{Univ. Lyon, ENS de Lyon, CNRS, Laboratoire de Chimie, F-69342, Lyon, France}

\author{Jérémie Margueritat}
\affiliation{Univ Lyon, Université Claude Bernard Lyon 1, CNRS, Institut Lumière Matière, F-69622, Villeurbanne, France}

\author{Benoit Mahler}
    \affiliation{Univ Lyon, Université Claude Bernard Lyon 1, CNRS, Institut Lumière Matière, F-69622, Villeurbanne, France}

\author{Asaph Widmer-Cooper}
    \affiliation{ARC Centre of Excellence in Exciton Science, School of Chemistry, University of Sydney, Sydney, New South Wales 2006, Australia}
    \affiliation{The University of Sydney Nano Institute, University of Sydney, Sydney,
New South Wales 2006, Australia}
    \email{asaph.widmer-cooper@sydney.edu.au}

\author{Benjamin Abécassis}
    \affiliation{Univ. Lyon, ENS de Lyon, CNRS, Laboratoire de Chimie, F-69342, Lyon, France}
    \email{benjamin.abecassis@ens-lyon.fr}

\date{\today}% It is always \today, today,

\begin{abstract}
The ability of thin materials to shape-shift is a common occurrence that leads
to dynamic pattern formation and function in natural and man-made structures.
However, harnessing this concept to rationally design inorganic structures at
the nanoscale has remained far from reach due to a lack of fundamental
understanding of the essential physical components. Here, we show that the
interaction between organic ligands and the nanocrystal surface is responsible
for the full range of chiral shapes seen in colloidal nanoplatelets. The
adsorption of ligands results in incompatible curvatures on the top and bottom
surfaces of the NPL, causing them to deform into helicoïds, helical ribbons, or
tubes depending on the lateral dimensions and crystallographic orientation of
the NPL. We demonstrate that nanoplatelets belong to the broad class of
geometrically frustrated assemblies and exhibit one of their hallmark features:
a transition between helicoïds and helical ribbons at a critical width. The
effective curvature $\bar{\kappa}$ is the single aggregate parameter that
encodes the details of the ligand/surface interaction, determining the
nanoplatelets' geometry for a given width and crystallographic orientation. The
conceptual framework described here will aid the rational design of dynamic,
chiral nanostructures with high fundamental and practical relevance.
\end{abstract}


\maketitle

Helical structures are fascinating chiral motifs that are found at different
length scales in Nature. Ranging from the molecular scale of DNA and proteins to
large-scale structures like tornadoes or galaxies, helices can be stabilised by
a broad range of driving forces. Beyond their aesthetic appeal, they can impart
useful properties, which has inspired the design of novel chiral materials with
specific mechanical or optoelectronic characteristics. At the nanometre scale,
helical shapes are especially coveted \cite{jiang_Emergence_2020} as their
chirality can endow them with desirable properties including chirality-induced
spin selectivity, circularly polarized luminescence or circular dichroïsm. In
this realm, nanoplatelets (NPL), a class of ultrathin 2D nanoparticles coated
with a monolayer of surfactants \cite{diroll_2d_2023}, are particularly relevant
because they display both outstanding optical properties and can be deformed
into helical shapes and assemblies. Aside from flat and tubular shapes, two
kinds of helices have been observed in NPL: purely twisted helicoids with
non-zero gaussian curvature and a straight centerline and helical ribbons which
have a cylindrical curvature and a helical centerline
\cite{guillemeney_curvature_2022}. The underlying physical mechanism which gives
rise to the emergence of this polymorphism is still unknown. It is clear from
previous work that ligands play a significant role in controlling NPL shape,
since ligand exchange can induce large shape variations in CdSe and CdTe NPL
\cite{jana_ligand-induced_2017,po_chiral_2022,Dufour_halide_2019,vasiliev_spontaneous_2018}.
However, the interplay between the NPL width, thickness, pitch, radius of
curvature, crystallographic structure/orientation, and the molecular
interactions at and near the ligand/crystal interface needs further
rationalization for predictive design to be possible. It is not yet clear which
of these parameters are most relevant for shape selection and hence which knob
one has to turn to achieve the desired structure. Moreover, the relevant
parameters which trigger the transitions between these different shapes have not
been identified, which calls for a unifying conceptual framework that can
rationalize the complete design space.

We provide here such a framework that comprehensively explains the polymorphism
of NPL and the emergence of chirality in these systems. The coexistence of
tubes, helicoïds and helical ribbons in NPLs leads us to hypothesize that they
belong to geometrically frustrated systems in which the bending and stretching
energies compete to dictate the shape of thin ribbons along with geometrical
constraints. Diverse systems at different length scales, such as surfactants or
chiral seed pods obey such a formalism \cite{ghafouri_helicoid_2005, selinger_shape_2004,grason_perspective_2016, hall_Building_2023a}, with the
common physical ingredients being a spontaneous curvature of microscopic origin
and a slender geometry. Using extensive molecular dynamics simulations, along
with theory and experiments, we show that the spontaneous curvature comes from
interaction between the ligands and the NPL crystal surface, which
induces directionally anisotropic stresses on the top and bottom surfaces and, hence, preferred curvature along different
crystallographic axes. Chirality results from the angle between the edge of the
NPL and the directions of preferred curvature, while the NPL's specific shape is dictated
by the scaling of the bending and stretching energies with respect to its width
and thickness.

\begin{figure*}
\begin{center}
\includegraphics[width=0.9\textwidth]{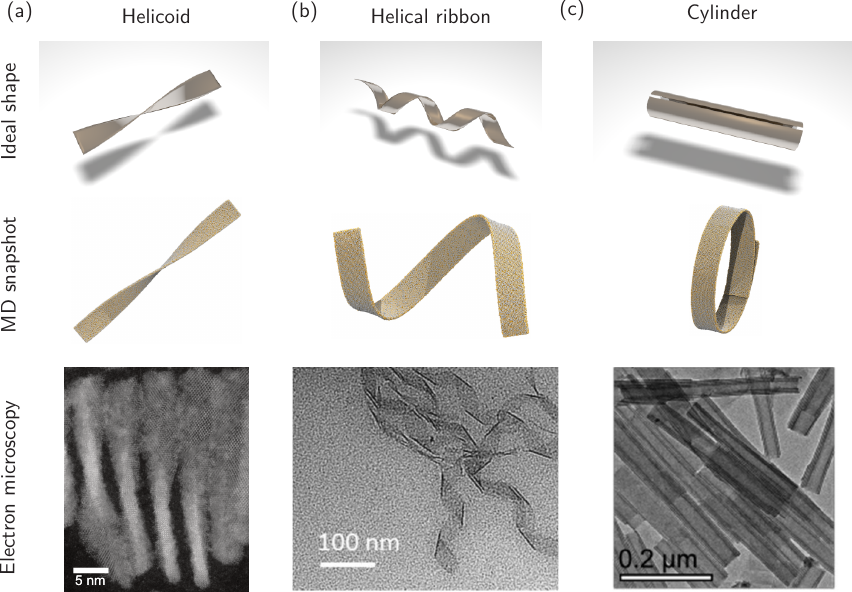}
\caption{CdSe colloidal nanoplatelet polymorphism. Three different shapes have
    been observed for CdSe NPL: (a) helicoids with a non-zero Gaussian curvature
    and a straight centerline, (b) helical ribbons with a helical centerline and
    a zero Gaussian curvature, and (c) tubes with a circular centerline and a
    zero Gaussian curvature. For each case, an idealized shape (top), a snapshot
    from an MD simulation (center) and an electron microscopy image (bottom) are
    shown. The MD simulation snapshots depict 3ML NPLs with a thickness of 1.1 nm and lateral dimensions of 150 nm $\times$ 10 nm. The TEM images are adapted from Refs.~\cite{jana_ligand-induced_2017},
    \cite{liu_photoluminescent_2019}, and \cite{Bouet_2dgrowth_2013}.}
\label{fig:polycryst}
\end{center}
\end{figure*}

\section*{Polymorphism of CdSe nanoplatelets}

We focus on zinc blende CdSe NPL \cite{Ithurria2008,*Ithurria2011} coated, in
their native state by carboxylate groups. NPL can display different geometries
depending on their lateral dimensions, thickness, and surface functionality, as
illustrated in Figure \ref{fig:polycryst}.  NPL with lateral dimensions from
\SIrange{5}{15}{\nano\meter} can either be flat or display a twisted helicoidal
shape with a non-zero Gaussian curvature in which the centreline corresponds to
the [100] crystallographic axis. Tube-like geometries with a mean radius of
curvature from \SIrange{10}{20}{\nano\meter} are observed when the NPLs have a
larger lateral dimension (hundreds of nanometers). Several reports show tubular
NPL with principal curvatures along either the [110] or [1$\overline{1}$0] axes
\cite{Bouet_2dgrowth_2013,kurtina_atomically_2019, huang_spontaneous_2019}.
Finally, helical ribbons have been reported
\cite{hutter_conformal_2014,liu_photoluminescent_2019}. In this geometry, the
Gaussian curvature is also zero, but the centreline is a helix that rotates
along the [110] axis with a defined pitch (e.g., $p$ = \SI{25}{\nano\meter}
\cite{hutter_conformal_2014} and $p$ = \SI{122}{\nano\meter}\cite{liu_photoluminescent_2019}).
This geometry is observed for intermediate widths ($w$ = \SI{27}{\nano\meter}
\cite{liu_photoluminescent_2019}) when one of the lateral dimensions is much
longer than the other. We thus observe a variety of shapes (helicoid, helical
ribbons, and tubes) which depend on the NPL's width and the crystal lattice's orientation with respect to the edges. Furthermore, several reports
provide evidence of shape changes when the surface chemistry of the NPL is
modified \textit{via} ligand exchange
\cite{Bouet_2dgrowth_2013,vasiliev_spontaneous_2018,kurtina_atomically_2019,
Dufour_halide_2019, po_chiral_2022}. This leads us to hypothesize that the
interaction between the surface of the NPL and the organic ligands is critical
in determining the shape of the NPL.

\begin{figure*}[htbp]
	\centering
	\includegraphics[width=0.9\textwidth]{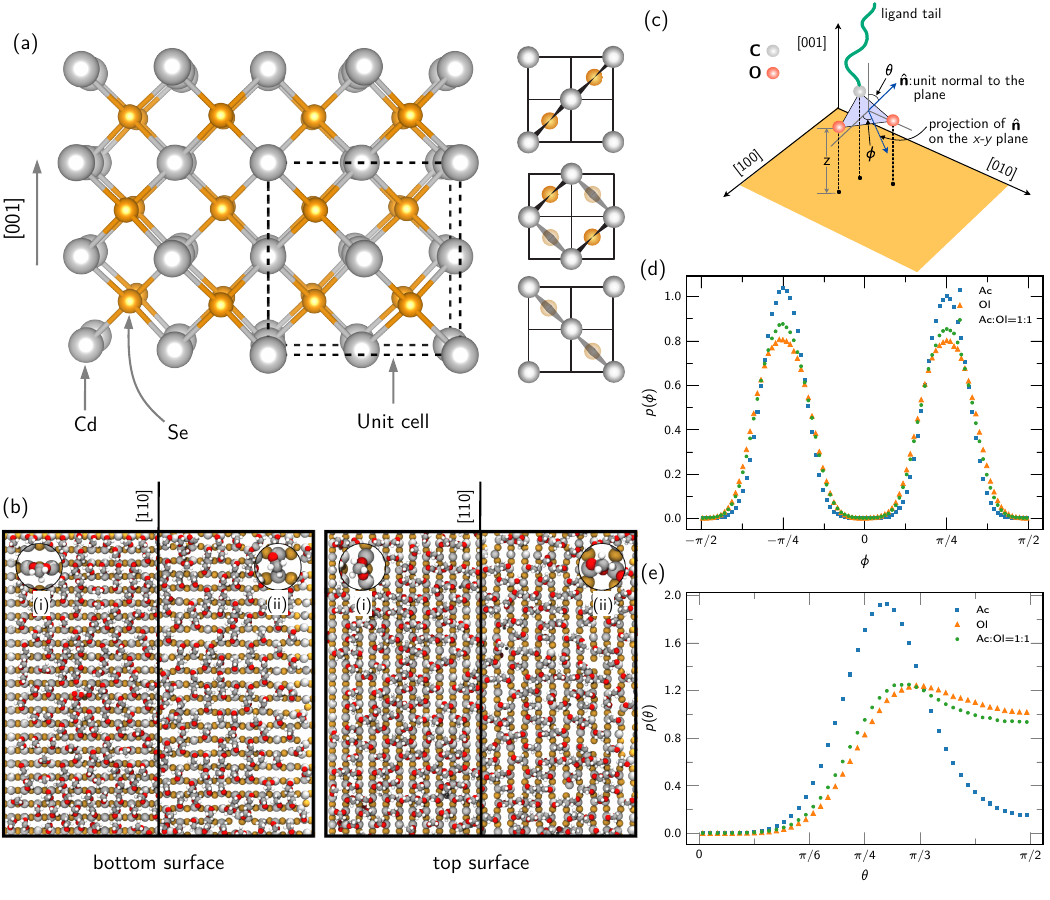}
    \caption{Ligand interaction with the NPL surface. (a) Crystallographic
      structure of CdSe zinc-blende NPL. The top and bottom facets correspond to
      the (001) planes. (b) View along the [001] axis showing  how Cd and Se atoms in adjacent layers are bonded. (c) Simulation snapshots of acetate ligands binding
      onto the top and bottom surfaces of a CdSe NPL (1.1 nm thick) along the [001]
      direction. (d) Schematic of the COO$^-$ binding moiety of acetate and
      oleate ligands on the surface of a CdSe NPL. The vector $\hat{\VC{n}}$ is
      the unit normal to the plane formed by the carbon and two oxygen atoms,
      which makes an angle $\theta$ with the $\VC{z}$-axis and whose projection
      on the $xy$-plane makes an angle $\phi$ with the $\VC{x}$-axis. The axes
      $\VC{x}$, $\VC{y}$, and $\VC{z}$ represent the [100], [010], and [001]
      crystallographic directions. $z$ is the perpendicular distance of an atom of
      the binding moiety from the $xy$-plane. Probability density functions of the
      angles $\phi$ (e) and $\theta$ (f) associated with the normal to the plane
      formed by the atoms of COO$^-$ for different ligands coating a CdSe NPL. The
      labels denote the following: Ac -- acetate coated NPL; Ol -- oleate coated
      NPL; AcOl -- 1:1 mixture of acetate and oleate.}
	\label{fig:ligconfig}
\end{figure*}

\section*{Ligand Induced Incompatible Curvatures}

To better understand the key ingredients of the ligand / NPL system, we
performed an extensive set of molecular dynamics (MD) simulations. We considered
zinc-blende CdSe NPL whose top and bottom facets are cadmium rich (001)
crystallographic planes coated with acetate and/or oleate ligands (see Methods
and  SI for details). Along its thickness, an N monolayer (ML) NPL presents N+1
cadmium planes and N selenium planes, corresponding to a thickness $t$
between \SI{0.6}{\nano\meter} (for 2ML) and \SI{3.5}{\nano\meter} (for 11ML).
We first simulated square 3ML CdSe NPL (t = \SI{1.1}{\nano\meter}, L = 10 nm)
with sufficient acetate ligands to provide charge neutrality
(\SI{3.7}{\text{ligands}\per\nano\meter^2}, close to the experimentally measured
surface density of
\SI{4.8}{\text{ligands}\per\nano\meter^2})\cite{singh_colloidal_2018}. Ligands
are first adsorbed on the flat NPL surface while the crystalline configuration
of the NPL remains fixed. Figure \ref{fig:ligconfig}c shows a resulting
configuration of ligands on the top and bottom surfaces of the NPL before (left)
and immediately (\SI{1}{\pico\s}) after (right) relaxation of the crystal. For
all of the monolayer compositions studied (acetate, oleate, and a 1:1 mixture of
the two), the azimuthal distribution of the ligands is strongly peaked at
\ang{45} and -\ang{45} (Figure \ref{fig:ligconfig}e), meaning that most of the
carboxylate moieties are oriented with their O-O axis perpendicular or parallel
to the [110] axis of the crystal (Figure \ref{fig:ligconfig}d). The ligand tail
length influences their polar orientation (Figure \ref{fig:ligconfig}f), with
acetate ligands having a peaked $p(\theta)$ distribution. At the same time,
monolayers that include oleic acid chains display a long tail distribution with
a significant probability for their tail to be perpendicular to the surface. We
found that acetate ligands bind to the Cd-rich (001) surfaces in two ways, shown
in Figure \ref{fig:ligconfig}c. In the bridging mode (i), the acetate molecule
stands symmetrically on top of two Cd atoms. In contrast, in the tilting mode
(ii), the acetate straddles a Cd atom asymmetrically, with one of the oxygens on
top of the Cd atom and the other in the space between two Cd atoms. These two
positions have preferred orientations: the first is observed along the dense
Cd-Se-Cd rows (which are perpendicular to each other on the top and bottom facets, see
Figure \ref{fig:ligconfig}b). In contrast, the second position generally happens
perpendicular to the dense rows. The dominant binding mode in our simulations is
the tilting mode (ii), oriented perpendicular to the dense rows. This is also
the preferred tilting mode predicted by DFT \cite{Koster_acetate_2016,
zhang_identification_2019}, though Ref.~\cite{zhang_identification_2019} argues
that a chelating mode is globally preferred. As discussed below, the direction
in which the NPL bends appears to be inherent to the crystal structure, with the
overall shape determined by the crystal orientation and dimensions and the
ligand type.

\begin{figure*}
	\centering
	\includegraphics[width=0.9\textwidth]{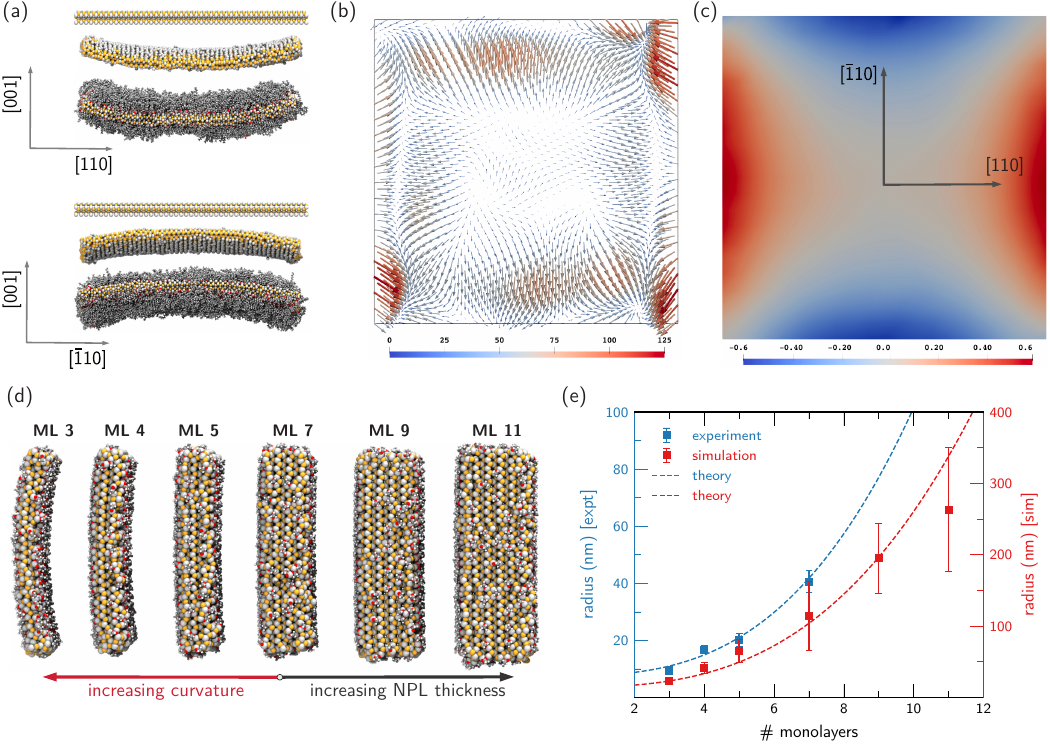}
  \caption{
  (a) Simulation snapshots of a square 3 ML CdSe NPL with \SI{10}{\nm} edge
  length and thickness 1.1 nm. The two viewing directions illustrate the
  non-zero Gaussian curvature.  (b) Spatial variation of in-plane forces (in pN)
  on the atoms of the midplane calculated by averaging over \SI{100}{\fs} to
  \SI{300}{\fs} after the commencement of deformation. The colors of the arrows
  indicate the magnitude of the forces. (c) Equilibrium displacement field of
  the NPL midplane (in nm) along the [001] direction. (d) Simulation snapshots
  for square NPL of increasing thickness (3 -- 11 ML, i.e. 1.1 nm -- 3.5 nm in
  thickness) from left to right. (e) The radius of curvature (nm) as a function
  of the number of monolayers in experiments and MD simulations. Experimental
  points come from Po \textit{et al.} \cite{po_chiral_2022}. The dashed lines
  represent fits to Eq.~9 of the SI Appendix with fitted parameters $\Psi
  \kappa_0=0.1328$ for experiments and 0.0712 for simulations and $e=
  \SI{2.752}{\nm}$ for experiments and \SI{2.456}{\nm} for simulations.  }
  \label{fig:squaremaps}
\end{figure*}

\section*{Elastic Theory of Thin Sheets}

When the ligands reach their equilibrium configuration, the NPL crystal is
relaxed yielding insight into how it deforms under the effect of surface
stresses induced by the ligands. We first consider square NPL with a 10 nm width
and edges parallel to the [110] and [$\bar{1}$10] directions (Figure
\ref{fig:squaremaps}). The relaxed NPL displays a saddle shape, as can be seen
in the MD snapshot (Figure \ref{fig:squaremaps}a) and in the displacement field
in Figure \ref{fig:squaremaps}c. This characteristic shape has non-zero Gaussian
curvature with a positive principal curvature direction along [110] and a
negative principal curvature along [$\bar{1}$10]. Our simulations are thus
consistent with the experimental observation of preferred bending along the
[110] directions. To understand the origin of this deformation, we mapped the
in-plane forces on the crystal atoms averaged over \SI{100}{\fs} to
\SI{300}{\fs} after the beginning of the crystal deformation (Figure
\ref{fig:squaremaps}b). We track down the origin of the deformation to
ligand-induced stresses on the top and bottom surfaces, which are
directionally anisotropic (Figure S1). The anisotropy in
stress can also be seen in Figure S3, particularly in the case of the
midplane. On the top surface, the anisotropic stress tends to bend the crystal
in the [110] direction hence imposing a preferred positive curvature, while on
the bottom surface, the stress direction is orthogonal and thus yields a
preferred curvature along the [$\bar{1}$10] axis. Due to surface stress, a
strain gradient appears along the [001] axis of the NPL, generating a
spontaneous curvature. 

The directional stress anisotropy between the top and
bottom layers is due to the interplay between the crystalline structure of the NPL and the cutting planes at their top and bottom. The NPL synthesis involves carboxylates, X-type ligands that bind preferably to cations \cite{Owen2015}. Hence, the top and bottom surfaces are cadmium-rich [001] surfaces. The improper rotation symmetry of the $\bar{4}$ axis in
the zinc blende ($\bar{4}3m$) structure means that as we go from the top surface to the bottom, the Cd-Se bonds rotate by 90° for each monolayer added (Figure \ref{fig:ligconfig}b). These bonds thus adopt a helical configuration whose axis is oriented along the [001] direction. In the end, the Cd-Se bonds at the top and bottom of the NPL are always oriented orthogonal to each other. Chirality arises when the edges of the NPL are misaligned with the directions of preferred curvature, inducing twisting, and one edge is longer than the other, such that one handedness of twisting has lower energy than the other. This is further discussed in the next section. Note that the mid-plane of the NPL is not a mirror symmetry plane since the reflection of a Se atom in the bottom
part corresponds to a vacant tetrahedral site in the top part. There are also no odd-even effects in the number of monolayers, with the preferred curvatures always orthogonal, whatever the number of monolayers, as shown in detail in Figure S4. This is consistent with twisted geometries observed for 3, 4 and 5 ML CdSe NPLs \cite{guillemeney_curvature_2022, kim_stacking_2019}.

Equilibrium configurations of NPL thus result from the minimization of elastic
energy, with curvatures imposed at the top and bottom layers by the surfactant
monolayer. This elasto-geometrical problem can be solved using the incompatible
elasticity theory of thin sheets, in which the total elastic energy is described
as the sum of bending and stretching energies. The two terms can be expressed as
a function of metric and curvature tensors and the material's mechanical
properties. This framework has proven accurate in describing a wide variety of
microscopic and macroscopic systems ranging from surfactant assemblies
\cite{Zhang2019} to chiral seed pods \cite{armon_geometry_2011}. In our case,
the ligands dictate a curvature only at the top and bottom surfaces (Figure
S1), while strains beyond the first atomic layers result
from strain propagation along the NPL thickness towards the mid-plane. We
developed a multi-layer model in the incompatible elasticity framework, which
features these boundary conditions (see theory section in the SI Appendix). The derived energy
functional shows that NPL can be described as ribbons with a pure spontaneous
twist, where the effective curvature is given by
\begin{equation}
    \bar{\kappa}=\kappa_0\Psi \frac{\Phi^3}{1+\Phi^3},
\label{eq:kappa_bar}
\end{equation}
where $\kappa_0$ is the curvature at the NPL/ligand interface,
$\Psi=Y_{\mathrm{lig}} / Y_{\mathrm{bulk}}$ is the ratio between the interfacial
and bulk Young's moduli, and $\Phi=e/t$ is the ratio between the effective
thickness of the surfactant layer and the NPL thickness.

\section*{Comparison with Simulation Results}

To test the scaling between the effective curvature and the thickness, we
simulated \SI{10}{\nano\meter} by \SI{10}{\nano\meter} square NPL and calculated
their radii of curvature along the [110] and [$\bar{1}$10] directions as a
function of NPL thickness (for acetate ligands). Results from these simulations,
as well as similar experiments \cite{po_chiral_2022}, are shown in
Figure~\ref{fig:squaremaps}e. While there are numerical differences between the
two, this is not surprising given the experiments were done on much larger NPL
that lie in the stretching-dominated regime and form tubes with cylindrical
curvature rather than saddle shapes. Fitting the variation of curvature with
thickness to Equation \ref{eq:kappa_bar} with two unknown parameters yields an
excellent fit with very similar values for the effective thickness of the
surfactant layer $e$, which is, as expected, of the same order of magnitude as
the thickness of the NPL. The larger deviation observed for the $\kappa_0\Psi$
prefactor between experiments and simulations can be explained by the difference
in experimental and simulation size regimes, as noted above. Other possible
contributions could be the accuracy of the force field used to describe the
ligand-surface interactions or uneven ligand coverage between the NPL surfaces
in experiments \cite{Lebedev2023}. Still, the overall agreement between theory,
experiment and MD simulations proves the relevance of our model in grasping the
key physical ingredients at play.

\begin{figure*}[htbp]
	\centering
	\includegraphics[width=0.95\textwidth]{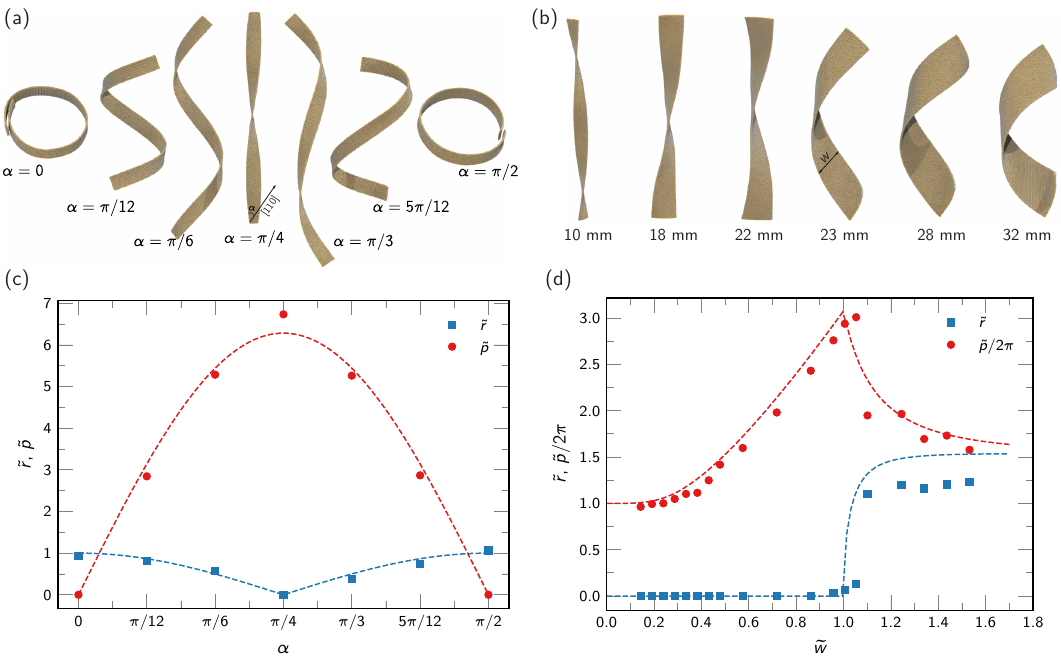}
  \caption{(a) Snapshots of 3ML CdSe NPL conformations from MD simulations as a
      function of the angle $\alpha$ between the [110] crystallographic
      direction and the long edge of the NPL. The NPL are 200 nm long (except
      for $\alpha=\pi/4$, which is 150 nm) and 10 nm wide with a thickness of
      1.1 nm, coated with a 1:1 mixture of acetate and oleate ligands (not shown
      for clarity). (b) Radius and pitch of the NPLs in (a) as a function of the
      angle $\alpha$, along with theoretical predictions (dashed lines) from
      Armon et al.~\cite{Armon2011} (c) Snapshots of CdSe NPL conformations from
      MD simulations for different widths at a fixed angle $\alpha = 45^{\circ}$
      between the [110] crystallographic direction and the long edge of the NPL.
      The NPLs are 150 nm long with a thickness of 1.1 nm and coated with
      acetate ligands (not shown for clarity).  (d) Radius and pitch for NPL
      similar to those in (c) as a function of width at $\alpha = \pi/4$, along
      with theoretical predictions (dashed lines) from Grossman et
      al.~\cite{Grossman2016} (see Sec. 1.4 for details).}
	\label{fig:angle_width}
\end{figure*}

An important parameter for shape selection in incompatible elastic ribbons is
the angle between the principal direction of curvature and the ribbon's
longitudinal axis. Since we previously showed that ligands induce principal
curvatures in the [110] and [$\bar{1}$10] directions, we define $\alpha$ as the
angle between the long edge of the NPL and the [110] axis.
Figure~\ref{fig:angle_width}a shows simulation snapshots for $\alpha$ varying
from 0 to $\pi/2$ for NPL with a ribbon geometry ($L_x = 10$ nm and $L_y = 100$
nm) covered in a 1:1 mixture of oleate and acetate ligands. The helicoid is the
favoured geometry for $\alpha= \pi/4$. As the angle increases towards $\pi/2$
(or decreases towards 0), the shape unwinds to finally reach a circular
tube-like shape for $\alpha=\pi/2$ (or 0). For chiral shapes, corresponding to
$\alpha \neq \pi/2$ or 0, the two enantiomers are obtained for $\pi/2 - \alpha$
and $\pi/2+\alpha$. We stress that shapes corresponding to $\pi/4 - \alpha$ and
$\pi/4+\alpha$ are not enantiomorphs since one can be obtained from the other by
a simple rotation. From the equilibrated shapes, we extract the pitch $p$ and
radius $r$ (see Methods) and report them as a function of $\alpha$ in Figure
\ref{fig:angle_width}c. For narrow strips, i.e., in the limit where the width $w
\rightarrow 0$, Armon \textit{et al.}\cite{Armon2011} derived the relations
$\tilde{r} = \cos 2\alpha$ and $\tilde{p} = 2\pi\sin 2\alpha$, where
$\tilde{r}=\bar{\kappa} r$ and $\tilde{p}=\bar{\kappa} p$ are the reduced radius
and pitch, respectively. Fitting the extracted values from the simulation to
these relations yields an excellent fit with the only free parameter
$\bar{\kappa} = 3.35 \times 10^{-2}$ nm$^{-1}$.

The hallmark feature of ribbons bearing incompatible curvatures is a
second-order phase transition between a helicoidal shape for small widths into a
helical shape as the width increases. The stretching energy scales as $w^5$
whereas the bending energy scales like $w$. Hence, when the width reaches a
critical value, it becomes energetically favorable to unwind the helicoid and
transition to a helical shape (or spiral) where the Gaussian curvature is
expelled. In this geometry, the stretching energy term dominates and NPL behave
as ribbons with a single intrinsic curvature that would be obtained by
stretching only one side (top or bottom). Figure~\ref{fig:angle_width}b shows
the shape evolution with increasing width $w$ at $\alpha = \pi/4$ for NPL with a
ribbon geometry ($L_y = 150$ \SI{}{\nano\meter}) covered with acetate ligands.
We observe a clear transition between helicoïdal ribbons at low $w$ and helical
ribbons as the width increases. From the simulated shapes, we extract the pitch
$p$ and radius $r$ (see Methods) and report them as a function of the reduced
width $\tilde{w} = w/w_c$ in Figure \ref{fig:angle_width}d. The unwinding
transition occurs for a critical width $w_c \approx 23$ nm for $\nu = 0.35$, $t
= 1.1$ nm with $\bar{\kappa} = 4.65 \times 10^{-2}$ nm$^{-1}$ as the only
fitting parameter for $\Tilde{r}$ and $\Tilde{p}$ as a function of $\Tilde{w}$
to equations in \cite{Grossman2016}. The small difference in $\bar{\kappa}$
between the two figures points towards the role of the ligand tails on the
reference curvature with pure acetate ligands allowing a larger $\bar{\kappa}$
than mixed oleate/acetate. Furthermore, we retrieve in simulations the
asymptotic behaviours predicted by theory: in the narrow limit the pitch
$\tilde{p}^{\dag} = 2\pi$, and in the wide limit the pitch $\tilde{p}^{\ddag} =
2\pi/\left(1-\nu\right)$ and the radius $\tilde{r}^{\ddag} =
1/\left(1-\nu\right)$.

\section*{Effective curvature}

The fact that all the simulated data can be fitted with only one free parameter
$\bar{\kappa}$, both for variations in $\alpha$ and $w$, is remarkable. This
unique parameter encodes the interaction between the ligands and the crystal
surface. Different ligand head-groups and binding motifs affect the imposed
curvature at the interface as shown experimentally for CdSe
\cite{Dufour_halide_2019} and CdTe \cite{vasiliev_spontaneous_2018}. This effect
is also observed in our MD simulations, where 5ML NPL coated with butanethiol (see Figure S2)
displays a much smaller radius of curvature (28 nm) than the same NPL coated
with acetate (77 nm, see Figure~\ref{fig:squaremaps}e). The greater curvature induced by the thiol ligands in our
simulations is consistent with the experimental observation that flat 5-6 ML
CdTe NPL roll up when their oleic acid ligands are replaced with
hexadecanethiol \cite{vasiliev_spontaneous_2018}.

More surprisingly, $\bar{\kappa}$ also depends on the interactions between the
ligand tails, which includes enthalpic contributions due to van der Waals
interactions and entropic effects related to the free volume available for the
ligand chains in the monolayer to undergo conformational changes. In simulations, the critical role of repulsive
entropic interactions between ligands is evidenced by curvature variations
observed when the $\epsilon$ parameter (which tunes the strength of vdW
attraction between the tails) is changed. For example, for 3 ML
\SI{20}{\nm}$\times$\SI{10}{\nm} CdSe NPLs at \SI{100}{\kelvin}, on reducing
$\epsilon$ from 100\%to 50\%, the radius of curvature of the NPL midline changes
from \SI{43.5}{\nm} to  \SI{36.9}{\nm} for octanoate ligands.  In addition, when
the chain length of linear thiolate ligands increases from 4 to 18 carbons, we
observe a signification decrease in curvature in MD simulations, in
semi-quantitative agreement with experiments (see Figure~S2).
Shapeshifting is even more dramatic when a branched ligand is used. Helical NPL
initially coated with carboxylates undergo complete unfolding when
functionalized with 2-ethyl-1-hexane thiol (see Figure~S5), showing
that entropic interactions between ligand tails play a key role in determining
NPL shape.

\section*{Conclusion and perspectives}
In this study, we demonstrated that the natural curvature of CdSe NPL is due to
interaction between ligands and the NPL crystal. This interaction induces
directionally anisotropic surfaces stresses, leading to preferred curvature
along different crystallographic axes on the top and bottom surfaces. The angle
between the NPL edge and these axes defines its chirality, while the balance of
bending and stretching energies relative to the NPL's width and thickness
determines its shape.

We traced back this effect's origin to the NPL's crystallographic structure.
Both top and bottom surfaces of the NPL are cadmium rich (001) planes and the
symmetry elements of the zinc-blende structure lead to orthogonally oriented
Cd-Se bonds on these surfaces. The shape of NPLs composed of other materials
could be explained by this theoretical framework if they exhibit similar
structural features. For example, other zinc-blende NPLs may exhibit frustrated
curvature effects \cite{vasiliev_spontaneous_2018}. More generally, we predict
that NPLs that have a crystal structure with a $\bar{6}$, $\bar{4}$ or $\bar{3}$
symmetry axis suitable oriented to their cutting planes have the potential to
exhibit such behavior. This prediction is consistent with a recent report of
chiral Gd$_2$O$_3$ NPLs \cite{liu_colloidal_2020}. Cubic Gd$_2$O$_3$ belongs to
the Ia3 space-group which has a $\bar{3}$ symmetry axis and, interestingly,
these NPL exhibit helical geometry in some instances. Chirality can also be
achieved for more symmetric crystal structures via the use of chiral ligands.
For example, gold has been shown to display twisted ribbon geometries when
functionalized with chiral thiols \cite{jiang_Emergence_2020}.

Another intriguing possibility is to use this new insight to design NPL that can
undergo large shape changes in solution with marginal changes in the structure
of the NPL/monolayer interface. Close to the unwinding transition ($w\simeq
w_c$), dramatic shape changes are expected to occur upon small $\bar{\kappa}$
changes. Since this parameter depends on the molecular details of the
self-assembled monolayer, we expect that changing the chemical identity of the
surfactants, e.g. via ligand exchange, could drive such a shape change. Another
feature of this class of systems is bistability \cite{Armon2014}, meaning that
two geometrically different conformations can have the same energy. It should thus
be possible to go from one configuration to the other with minimal energy input
to the system. This feature could be exploited in stimuli-responsive
nanomaterials by applying the theory and physical insights described in this
paper, linking the molecular structure of ligands with the morphology of
inorganic NPL.

\begin{acknowledgments}
This article is part of a project that has received funding from the European
Research Council (ERC) under the European Union’s Horizon 2020 research and
innovation program (Grant agreement No. 865995). Part of this work was supported
by the Australian Research Council through the Centre of Excellence in Exciton
Science (CE170100026). Computational resources were provided by the University
of Sydney HPC service and the Pawsey Supercomputing Centre with funding from the
Australian Government and the Government of Western Australia. We thank Efi
Efrati, Arezki Boudaoud, Yo\"el Forterre and Lyd\'eric Bocquet for stimulating
discussions. We thank Sandrine Ithurria for providing some data of
Ref.~\citenum{po_chiral_2022} before publication. This work was granted access
to the HPC resources of IDRIS under the allocation 2022-AD010913529 made by
GENCI. We also gratefully acknowledge support from the PSMN (P\^ole Scientifique
de Mod\'elisation Num\'erique) computing centre of ENS de Lyon.
\end{acknowledgments}

\section*{Materials and Methods}

\subsection{Synthesis and functionalisation of CdSe Nanoplatelets}
\subsubsection{Synthesis of 3ML CdSe NPLs}
In a typical synthesis, 140 mL ODE, 2.22 g Cd(Acetate)2·2H$_2$O and 2.23 mL oleic acid are introduced in a 250 mL three-neck flask. The temperature of the solution is raised to 200°C under Argon flux. After injection of 3.6 mL of TOPSe 1M, the reaction mixture is annealed at the same temperature for 10 minutes, forming 3 MLs CdSe NPLs. At the end of the reaction, 10 mL of oleic acid is added before the solution cools down to room temperature. NPLs are isolated from the reaction mixture by centrifugation; they are then dispersed in hexane and precipitated one more time by adding ethanol. The final product is dispersed in hexane.

\subsubsection{Functionalisation by thiols}
200 $\mu$L of the previous 3ML CdSe NPL dispersion are added to 1mL of toluene. 1 mmol (large excess amount) of the appropriate thiol is added, and the solution is heated for three days at 65°C. The NPLs are precipitated by adding ethanol, the dispersion is centrifuged at 9000 rpm for 3 minutes and the precipitate is redispersed in toluene. The purification step is repeated once.

\subsection{Molecular dynamics simulations}
\label{ss:md}
Molecular dynamics (MD) simulations of isolated ligand-coated CdSe NPL were
performed at constant volume and temperature (\SI{100}{\K} for thiol-coated NPL
and \SI{300}{\K} for the remaining systems), maintained via a Nos\'e-Hoover
thermostat, using the LAMMPS software package \cite{Thompson2022}. The NPL and
ligands were modeled with explicit atoms, while the solvent was treated
implicitly, in order to limit computational costs, by reducing the interaction
strength between ligand chains to 85\% (3ML oleate-coated NPL) and 50\% (5ML
thiol-coated NPL)  of their original value. These values were chosen based on
results from previous work with alkanethiols coating spherical nanoparticles
\cite{Kister2018} and on the fact that ligand chains remained disordered and
mobile at the chosen temperature. Following Rabani \cite{Rabani2002}, the
interaction potential for CdSe consists of a short-ranged 12-6 Lennard-Jones
potential combined with a long-range Coulombic contribution, with parameters
given in Ref.~\citenum{Rabani2002}. Ligand interactions were modeled using the
OPLS-AA force field \cite{Jorgensen1996}. The set of parameters for each type of
ligand were obtained from the LigParGen web server \cite{Jorgensen2005,
Dodda2017, Dodda2017a}. The MD equations of motion were integrated using the
velocity-Verlet algorithm with a time step of \SI{1}{\fs}.

CdSe NPL with Cd-terminated [001] polar surfaces (see Fig. \ref{fig:ligconfig}a)
were constructed by first placing the Cd and Se atoms on a zinc blende lattice
(see SI Appendix for calculation of the lattice parameter). The
structures thus obtained were cut by a set of parallel planes along the
crystallographic axes to obtain the desired sizes and the number of monolayers.
For NPL with different orientations of the [110] axis, the lattice was rotated
by appropriate angles prior to cutting. Next, the crystalline core was coated
with either acetate, oleate, or thiolate ligands at a density of 3.27 ligands
nm$^2$, which ensured charge neutrality of the NPL. The coating was performed by
randomly placing the ligands within a thin shell encompassing the core and then
optimizing their positions using the software package PACKMOL
\cite{Martinez2009}.

The resulting ligand-coated core was placed at the center of a simulation box,
with at least \SI{50}{\angstrom} of vacuum added on all sides. In almost all
cases, the simulation box was a rectangular parallelepiped, where the longest
and shortest dimensions were of the order of the associated NPL dimensions in
their undeformed state. This ensured that in practice, after equilibration, the
box dimensions remained significantly larger than the NPL size. Periodic
boundary conditions were applied along all three dimensions.

Equilibration was performed \textit{via} a sequence of stages. First, an energy
minimization run was carried out to remove any significant overlap of the atoms.
Second, the ligands were relaxed at constant energy with a maximum atomic
displacement of \SI{0.1}{\angstrom} per time step for a duration of
\SI{50}{\ps}. Third, the ligands were relaxed at the desired temperature for
another \SI{100}{\ps}. Note that during the above three stages, only the atoms
belonging to the ligands were displaced and the CdSe core was held static.
Finally, both the ligands and the crystalline core were allowed to equilibrate
together at the chosen temperature for \SI{6}{\ns}. Equilibration was followed
by a production run of 3-25 ns, depending on the NPL size and whether
significant fluctuations persisted in the resulting configurations. The
configurations were sampled every \SI{10}{\ps}. All results presented here have
been averaged over all the sampled configurations.

\subsection{Characterisation of bulk elastic constants}
\label{ss:bulk}
\begin{table*}[t]
\centering
\renewcommand{\arraystretch}{1}
\caption{Calculated lattice constants $a$ (in \AA), elastic constants $C_{ij}$ (in GPa), and bulk moduli $B$ (in GPa) of zincblende CdSe. Results obtained from Eran Rabani at 0 K are compared with our results at the same volume and temperature (NVT) and at 1 atm pressure (NPT). The experimental lattice constant is \SI{6.08}{\angstrom} \cite{Cohen1988}.}
\begin{tabular}{p{2.0cm}p{2.5cm}p{2.5cm}p{3.0cm}p{3.0cm}}
\hline
Property & Rabani (0 K) & NVT (0 K) & NPT (0 K, 1 atm) & NPT (300 K, 1 atm) \\ \hline
$a$ & 6.13 & 6.13 & 6.17 & 6.20 \\
$C_{11}$ & 53.8 & 53.6 & 53.3 & 53.2 \\
$C_{12}$ & 40.2 & 39.0 & 38.5 & 40.4 \\
$C_{44}$ & 24.1 & 24.0 & 23.1 & 38.1 \\
$B$ & 44.7 & 44.0 & 43.5 & 44.6 \\ \hline
\end{tabular}
\label{tab:elastic_constants}
\end{table*}

The deformation of CdSe NPL depends in part on the elastic properties of the underlying zincblende CdSe lattice, which has not been determined experimentally. For reference, we therefore report the bulk elastic constants for this lattice under several thermodynamic conditions using the same potential used to simulate the NPL (see Table~\ref{tab:elastic_constants}). These values were reported incorrectly in prior work,~\cite{Rabani2002,Gruenwald2012} so for comparison we list values provided directly to us by Eran Rabani.

The bulk elastic constants were determined by placing a cubic 3D crystal inside a periodic simulation box, letting it equilibrate, and then measuring changes in the stress tensor as the box was deformed. Three different cases were considered: a temperature of \SI{0}{\K} and constant volume (to match the lattice constant reported previously)~\cite{Gruenwald2012}; a temperature of \SI{0}{\K} and a pressure of 1 atm; and at a temperature of \SI{300}{\K} and a pressure of 1 atm. The temperature and pressure were maintained using a Nos\'{e}-Hoover thermostat and barostat, respectively.

\subsection{Characterization of NPL shape}
\label{ss:shape}
At equilibrium, ligand-coated CdSe NPL adopt deformed conformations that possess either a helicoidal midplane or a helical midline. Such shapes can be characterized by their radius and pitch. However, 
the calculation of these quantities directly from atomic positions is cumbersome and prone to excessive noise in the data. To alleviate these issues, we employed an approach based on an atom-to-continuum mapping to extract the midplane and the midline in the form of a surface mesh and line mesh, respectively. Our approach has been informed and influenced by several works in the solid mechanics community \cite{Zimmerman2004, Zimmerman2010, Templeton2010, Admal2010, Admal2011, Rigelesaiyin2018, Wang2021}, where the primary emphasis was on calculating inhomogeneous continuum fields, e.g., strain and stress, from molecular dynamics simulations.\\

We considered the undeformed state of an NPL as a reference configuration with respect to which any deformation will be calculated. Since we are interested in only the midplane and the midline, it was sufficient to consider only the crystalline CdSe core. We constructed a three dimensional grid spanning the region in space occupied by the Cd and Se atoms. To reduce the number of grid points as well as to improve the accuracy of the subsequent interpolations, the grid points were chosen to be Chebyshev-Gauss-Lobatto nodes, appropriately mapped from the unit domain $[-1,1] \times [-1,1] \times [-1,1]$. Atomic displacements were calculated, for each Cd and Se atom, from their positions in the reference and the deformed states. Next, the atomic displacements were projected onto the grid points via a localization function. The localization function $\psi$ is a regularized approximation with compact support to a $\delta$-function  that averages any atomic property over the neighboring atoms of a grid point.  Its functional form is that of a quartic spline (see Eq~(4.5) in  Ref.~\citenum{Admal2010}): 
\begin{equation}
    \psi = \begin{cases} 
        \frac{105}{16\pi r_w^3} \left(1 + \frac{3r}{r_w} \right) 
        \left(1 - \frac{r}{r_w} \right)^3 & \qquad \text{if}\ r \leq r_w, \\
         0 & \qquad \text{otherwise},
    \end{cases}
\end{equation}
where the support radius $r_w$ was taken equal to \SI{8}{\angstrom} for calculating the displacement field and \SI{20}{\angstrom} for the force and stress fields.  While the exact value of $r_w$ is subject to choice, a higher value will increase computational cost, and too low a value will result in poor spatial averaging. The chosen value of \SI{8}{\angstrom} is greater than the size of the CdSe unit cell, and varying it generated only minor changes in any subsequent calculations. However, atomic forces and stresses being highly fluctuating quantities both spatially and temporally, we had to increase $r_w$ to approximately three unit cells in order to obtain a smoother field. Note that $\psi$ is defined in terms of the coordinates of the reference configuration, hence after projecting the atomic displacements onto the reference grid the resulting continuum displacement field is a material field \cite{Zimmerman2010}. Furthermore, following Zimmerman et al \cite{Zimmerman2010}, we used a mass-weighted average when calculating all field quantities from the respective atomic quantities. \\

The displacement field was averaged over all NPL configurations sampled over a
simulation run, and then interpolated to the midplane and the midline. These
interpolations were performed onto a uniform grid in two and one dimensions,
respectively. The nodal coordinates of deformed state of the midplane and
midline were directly obtained by adding the nodal displacements to the nodal
positions in the reference state. From the nodal coordinates of the deformed
midplane, the nodal surface normals were calculated by averaging the normals to
all quadrilaterals incident onto a node.\\

The pitch of helicoidal midplanes was obtained by the kinematic surface fitting
method presented in Refs.~\citenum{Pottmann1998} and \citenum{Andrews2013}.
In this case, an NPL undergoes pure twisting, hence the radius is zero. In
addition to the nodal positions, the fitting algorithm also requires the nodal
surface normals. For NPL with a helical midline, their radii were determined by
fitting a cylinder to the coordinates of the midline using the orthogonal
distance regression method discussed
in Ref.~\citenum{Shakarji1998}. The fitting algorithm requires an initial
estimate for the the axial direction of the helix and the radius. The former was
determined from a combination of singular value decomposition of the midline
coordinates and visual inspection. Next, to obtain an estimate of the radius,
the coordinates were projected onto a
plane perpendicular to the axis and fit to a circle using Taubin's method
\cite{Taubin1991, Chernov2010} . The fit to a
cylinder yields the final values of the radius and the helix axis. Finally, the
pitch was obtained from the linear relation between the axial and the angular
displacements on traversal along the midline.\\

For square ligand-coated CdSe NPL that bent along the [110] and [$\bar{1}$10] directions adopting a saddle shape geometry, we calculated surface curvatures by fitting a circular arc to the inner layers of atoms of the crystals. We fitted a circle to the line of atoms of interest by employing an orthogonal least squares algorithm in which the sum of squares $ f = \sum^{n}_{i=1} {d_i}^2 $ is minimized. In this equation, $d_i$ is the geometric distance between the data point of choice ($x_i$, $y_i$) and the hypothetical circle, and is given by $ d_i = x_i-a_2 + y_i-b_2-R $, where $(a,b)$ is the center of the circle, and $R$ is its radius. We used Taubin’s method to minimize $f$ algebraically \cite{Taubin1991}, taking only the inner layers of atoms in a NPL into account to reduce error, as atoms directly in the surface of the crystal tend to be more disordered due to their strong interaction with ligands making quantification of the curvature more difficult. Additionally, we removed gross outliers if those were still present \cite{guo-curv-2018}.

\subsection{In-plane forces calculation}
The in-plane forces for Cd layers were averaged within each quadrant of the acetate-coated NPL (Figure \ref{fig:forces_quadrant}) over the interval between \SI{100}{\femto\s} to \SI{300}{\femto\s} following crystal relaxation in the NVT ensemble at \SI{100}{\kelvin}. This step was preceded by a \SI{1}{\pico\s}-long ligand relaxation stage, where ligands were allowed to find their preferred position on the surface of the fixed crystal in the absence of solvent.

\subsection{Theory}
We consider a ligand-coated NPL as a multilayered ribbon of length $L$ and width $w$. The ribbon consists
of a crystalline core of thickness $t$, bounded by two layers of ligands at the top and the bottom, each of
effective thickness $e$. The crystalline core is in turn composed of $N$ layers, each of uniform thickness $t_{\mathrm{single}}$, and thus $t = N t_{\mathrm{single}}$. We further assume that $L \gg w \gg (t+2e)$.
In a typical experiment, $e$ exceeds $t_{\mathrm{single}}$ by a factor of at least five.
At the  top and the bottom, a unidirectional curvature of magnitude $\kappa_0$ is induced by the interactions the ligands. The reference curvature $\bar{\VC{b}}_{\pm}$ at the edges is given by
\begin{equation}
    \bar{\VC{b}}_{\pm} = \pm \kappa_0
    \begin{pmatrix}
        \cos^{2}\alpha_{\pm} & \cos\alpha_{\pm} \sin \alpha_{\pm} \\
        \cos\alpha_{\pm}\sin \alpha_{\pm}  & \sin^2 \alpha_{\pm} 
    \end{pmatrix},
\end{equation}
where $\pm$ indicates the top ($+$) or bottom ($-$) layers and $\alpha_{\pm}$ are the directions of the curvature relative to the ribbon midline. Since the top and the bottom are the same up to an orientation, $\alpha_- = - \alpha_+$. Henceforth we will assume $\alpha_+ = \pi/4$. Thus
\begin{equation*}
    \bar{\VC{b}}_{\pm}  = \frac{\kappa_0}{2}
    \begin{pmatrix}
        \pm 1   &  1 \\
            1   & \pm 1
    \end{pmatrix}.
\end{equation*}
Without the edges, the  ribbon’s bulk energy density is given by \cite{Grossman2016}
\begin{equation}
\begin{split}
    E_{\mathrm{bulk}} 
    &= Y_{\mathrm{bulk}} \left[ \frac{t w^5}{80}  \left( ln - m^2 \right)^2 
        + \frac{t^3 w}{3} \Bigl\{ \left(l+n\right)^2 \Bigr . \right .\\
            &\qquad \left. \Bigl. -2\left( 1-\nu \right)\left( ln-m^2 \right)\Bigr\} \right] \\
    &=Y_{\mathrm{bulk}} \left[ \frac{t w^5}{80} \det \VC{b}^2 
        + \frac{t^3 w}{3} \Bigl\{ \Tr^2 \VC{b} \Bigr . \right .\\
           &\qquad \left. \Bigl. -2\left( 1-\nu \right) \det \VC{b}\Bigr\} \right],
\end{split}
\end{equation}
where $Y_{\mathrm{bulk}}$ is the Young's modulus of the bulk crystal, $\nu$ is the
Poisson's ratio, and $l$, $m$, $n$ are the normal curvature, twist, and binormal curvature, at
the midsurface, i.e. $\VC{b} = \begin{pmatrix}l & m \\ m & n \end{pmatrix} $.

Assuming the length scale of curvature to be much greater than that of the thickness, the curvatures at the top and bottom surfaces can be obtained from the curvature at the mid-surface as 
\begin{equation}
\begin{split}
    \VC{b}_{\pm} &= \VC{b} \mp \frac{t}{2} \VC{b}^2 \\
    &= \begin{pmatrix} l & m \\ m & n \end{pmatrix} \mp \frac{t}{2}
        \begin{pmatrix} l^2+m^2 & m\left(l+n\right) \\
            m\left(l+n\right) & m^2+n^2\end{pmatrix}.
\end{split}
\end{equation}
Thus the top and bottom surfaces of the ribbon contribute additional energy stemming from the difference between the curvature and the reference curvatures $\bar{\VC{b}}_{\pm}$.
As the thickness of the ligands is larger than that of a single layer the top and bottoms ribbons additional contribution to the energy density is given, to leading order by
\begin{equation}
\begin{split}
    \Delta E &= Y_{\mathrm{lig}} 
    \Biggl[ \frac{e w^5}{40} \left(ln-m^2\right)^2
        + \frac{e^3 w}{3} 
        \Bigl\{
            \Tr^2 \left(\VC{b}_{+}-\bar{\VC{b}}_{+}\right) \Bigr. \Biggr.\\
            &\quad - 2\left(1-\nu\right) \det \left(\VC{b}_{+}-\bar{\VC{b}}_{+}\right) 
                +\Tr^2 \left(\VC{b}_{-}-\bar{\VC{b}}_{-}\right) \\
            &\quad \Biggl.\Bigl. - 2\left(1-\nu\right) \det \left(\VC{b}_{-}-\bar{\VC{b}}_{-}\right)
        \Bigr\}
        \Biggr]
\end{split}
\end{equation}
where $Y_{\mathrm{lig}}$ is the Young’s modulus of the outer ribbons. 
Using  a little algebra, the additional term can be written, to
leading order (assuming the curvature is much greater compared to the thickness)
\begin{equation}
\begin{split}
    \Delta E & = 2Y_{\mathrm{lig}} 
    \Biggl[ \frac{e w^5}{80} \left(\det \VC{b} \right)^2
        + \frac{e^3 w}{3} \Bigl\{ \Tr^2\left(\VC{b}-\VC{B}\right)\\
        &  \quad -2\left(1-\nu\right) \det \left(\VC{b}-\VC{B}\right) + \left(1+\nu\right)\kappa_0^2 
        \Bigr\}
        \Biggr]
\end{split}
\end{equation}
where $\VC{B} = \begin{pmatrix}0 & \kappa_0\\ \kappa_0 & 0 \end{pmatrix}$. Note that the last term is a constant that may be discarded from the energy as it does not contribute to the mechanics. Finally, introducing the thickness ratio $\phi = e/t$ and the energy ratio $\psi = Y_{\mathrm{lig}}/Y_{\mathrm{bulk}}$ we may write the total effective energy density as
\begin{equation}\label{eq:rib_effective}
\begin{split}
    E &= E_{\mathrm{bulk}} + \Delta E \\
    &= Y_{\mathrm{bulk}} \Biggl[
        \frac{\left(1+\psi\phi\right)tw^5}{80} \left(\det \VC{b}\right)^2 \Biggr . \\
        &\quad \Biggl . + \frac{t^3w}{3} \Bigl\{\Tr^2\left(\VC{b}-\bar{\VC{b}}\right) 
            - 2\left(1-\nu\right) \det\left(\VC{b}-\bar{\VC{b}}\right) \Bigr\}
        \Biggr] \\
        & \quad + E_0
\end{split}
\end{equation}
where $\bar{b} = \frac{\psi \phi^3}{1+\phi^3}B$ is the effective reference curvature, and 
\begin{equation}
\begin{split}
E_0 &= 2Y_{\mathrm{lig}}
\frac{e^3}{3}w \left[ 
    \left(1+\nu\right)\kappa_0^2 \right .\\
    &\quad \left . + \frac{\phi^2}{1+\phi^3} \Bigl\{ \Tr^2 \VC{B} -2\left(1-\nu\right)\det \VC{B}\Bigr\}\right]
\end{split}
\end{equation}
is the residual energy that does not affect the mechanics.
The energy in \eqref{eq:rib_effective} has the same form as that of a general ribbon with residual stress \cite{Grossman2016}, which indicates that the system behaves as a ribbon with a spontaneous twist with an effective curvature $\bar{\kappa} = \psi \kappa_0 \phi^3/\left(1+\phi^3\right)$. The radius of curvature can
then be rewritten as:
\begin{equation}
    r_c =\frac{1}{\psi \kappa_0}\left( 1 + \left(\frac{t}{e} \right)^3 \right).
    \label{eq:radius_thickness}
\end{equation}
The critical width $w_c$ of a ribbon for helicoid-helix transition is 
\cite{Grossman2016, Zhang2019}
\begin{equation}
    w_c = \Biggl[ \frac{320}{3} \frac{1+\nu}{\left(1-\nu\right)^2} \Biggr]^{1/4} 
        \sqrt{\frac{t}{\bar{\kappa}}},
\end{equation}
Note that a similar formula given in  Ref.~\citenum{Grossman2018} (Eq.~6a) has the signs flipped in the
numerator and denominator due to a difference in geometry. Following earlier works \cite{Armon2011, Armon2014,
Grossman2016, Grossman2018, Zhang2019} we express the NPL width $w$, pitch $p$, and radius $r$ in their nondimensionalized form $\tilde{p} = p\bar{\kappa}$, $\tilde{r} = r\bar{\kappa}$, and $\tilde{w} = w/w_c$. For narrow strips, i.e., in the limit $\tilde{w} \rightarrow 0$, the radius and pitch are given by \cite{Armon2011} 
\begin{equation}
    \begin{gathered}
        \tilde{r} = \cos 2\alpha \\
        \tilde{p} = 2\pi\sin 2\alpha.
    \end{gathered}
\end{equation}
For wide strips with $\alpha = \pi/4$ the asymptotic values of the radius and pitch are \cite{Armon2011}
\begin{equation}
    \tilde{r} = \frac{\tilde{p}}{2\pi} = \frac{1}{1-\nu}.
\end{equation}

\bibliography{biblio}% Produces the bibliography via BibTeX.

\clearpage
\renewcommand{\thefigure}{S\arabic{figure}}
\setcounter{figure}{0}  

\section*{Supplementary Material}

\begin{figure*}[htbp]
	\centering
	\includegraphics[width=0.9\textwidth]{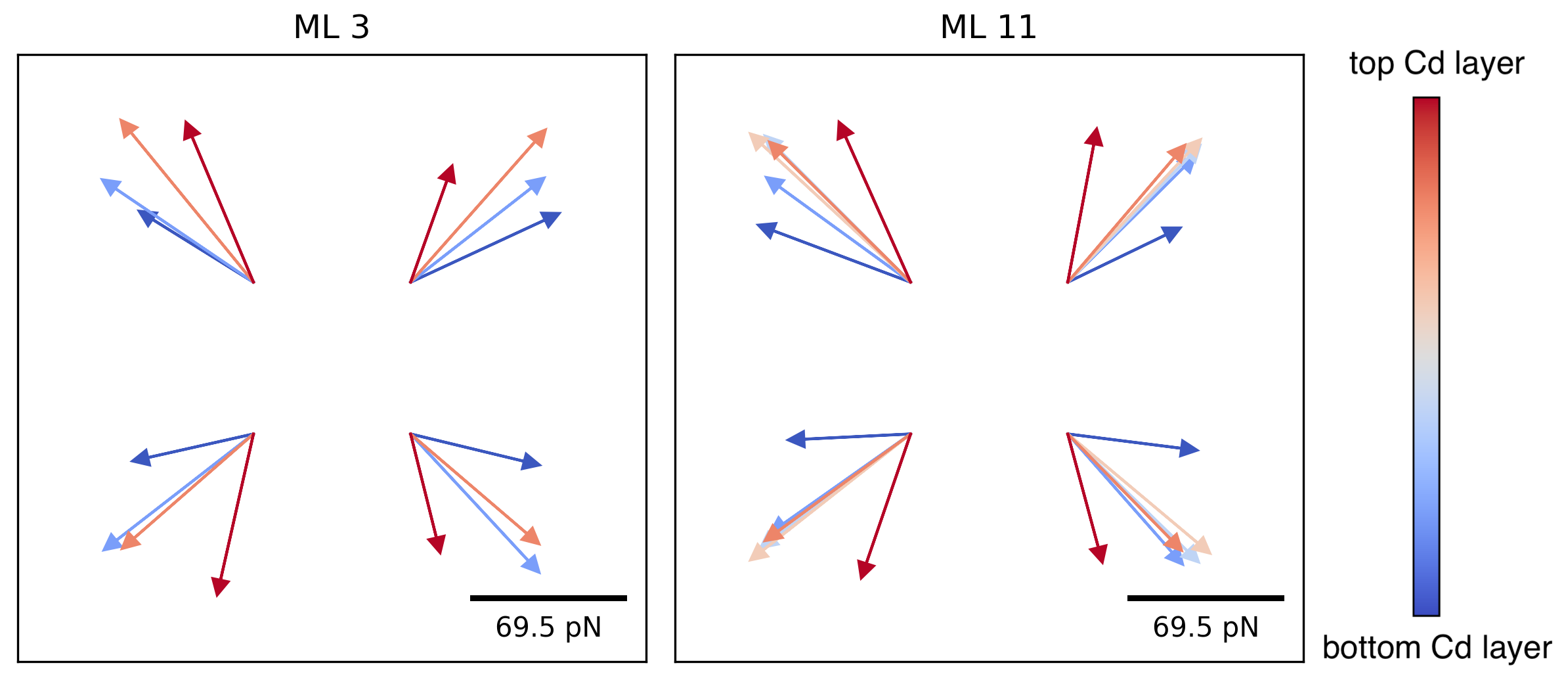}
    \caption{In-plane forces, averaged over each quadrant, for the Cd layers (which correspond to the (001) planes) of a 3 ML (left) and a 11 ML (right) \SI{10}{\nano\meter} by \SI{10}{\nano\meter} NPL with $\alpha=0$. The thickness of the 3ML NPL is 1.1 nm, and that of the 11 ML NPL is 3.5 nm. The scale bar on the right of the plots provides a measure of the magnitude of the forces, while the arrows show their direction. Colors represent different Cd layers, going from the nanocrystal's bottom (dark blue) to the top (dark red) surface.}
	\label{fig:forces_quadrant}
\end{figure*}

\begin{figure*}[htbp]
	\centering
	\includegraphics[width=0.95\textwidth]{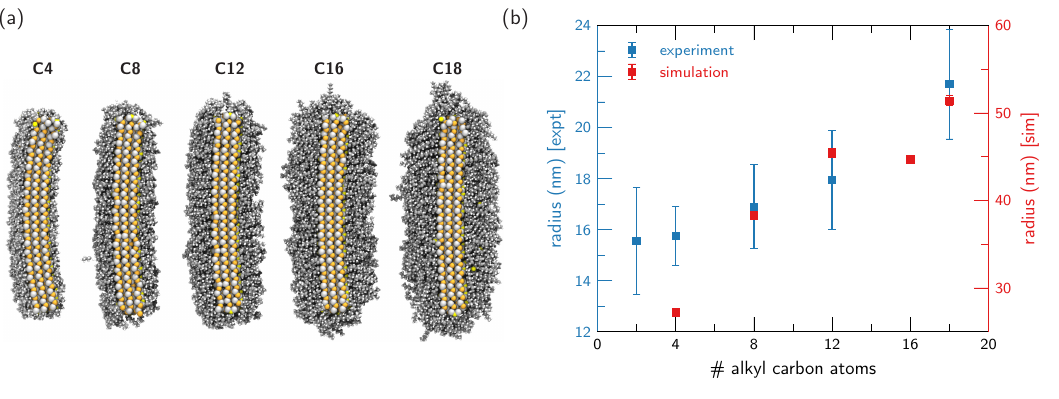}
    \caption{Effect of ligand chain length on the curvature. (a) Snapshots of simulated CdSe NPL of lateral dimensions 10 nm $\times$ 10 nm and thickness 1.7 nm (5 ML) at $\alpha=0$ coated with linear alkyl thiol ligands with 4-18 carbon atoms in length. (b) The radius of curvature as a function of carbon number from MD simulations and experiments by \cite{po_chiral_2022}. Error bars indicate 95\% confidence intervals. The numerical difference between the two data sets is partially due to the simulations being run at lower temperatures than the experiments (\SI{100}{\kelvin} vs \SI{300}{\kelvin}), which was done to obtain more reliable estimates for the radius of curvature.}
	\label{fig:ligand_effect}
\end{figure*}

\begin{figure*}[htbp]
	\centering
	\includegraphics[width=0.95\textwidth]{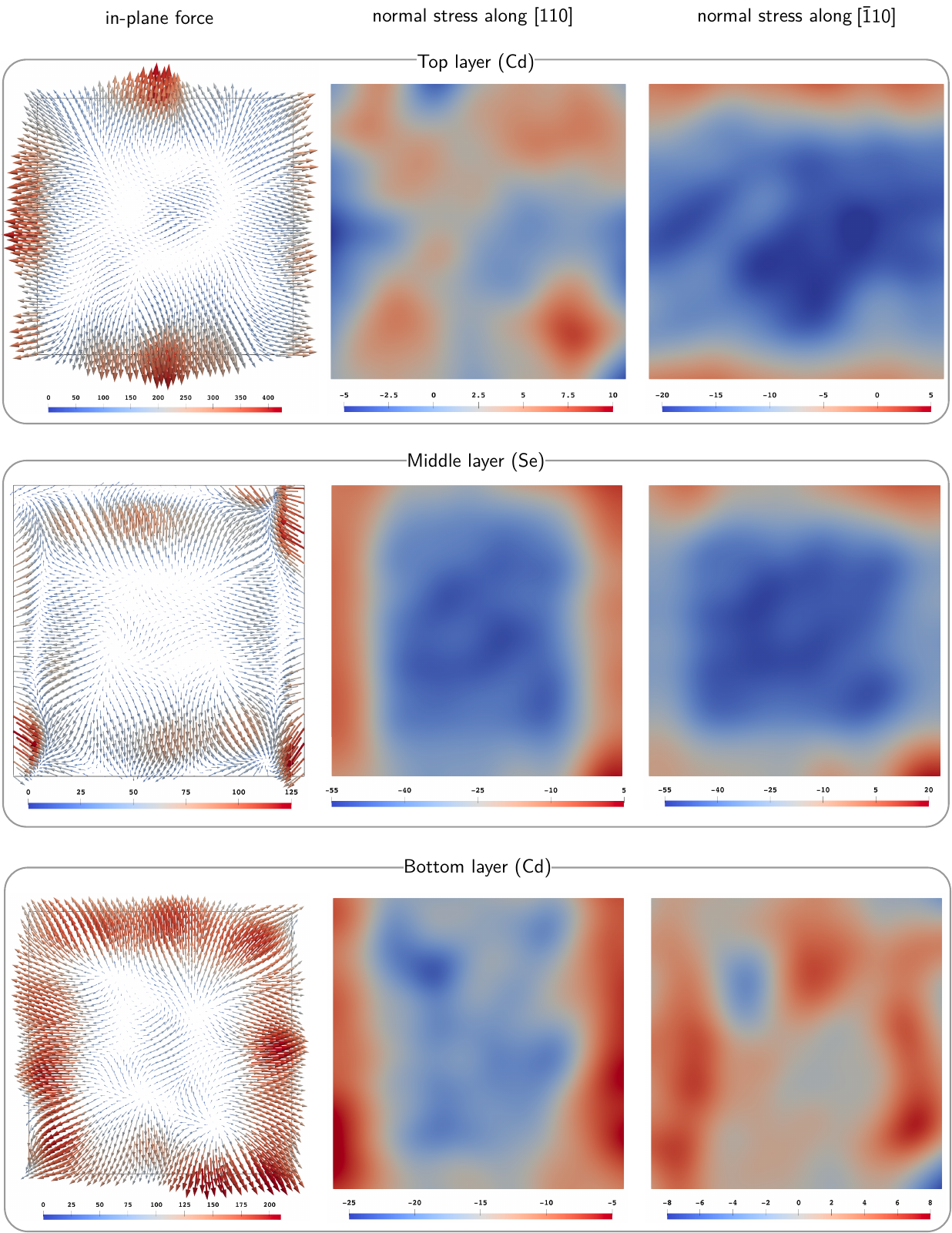}
    \caption{In-plane forces on each atom (in pN) and normal stresses (in GPa) at the top, middle, and bottom layers of a 3 ML (thickness 1.1 nm) \SI{10}{\nm} $\times$  \SI{10}{\nm} CdSe NPL coated with acetate ligands at \SI{100}{\kelvin}, obtained from time-averaging from \SI{100}{\fs} to \SI{300}{\fs} after the onset of deformation. In the case of the force plots, the color coding indicates the magnitude of the in-plane forces.}
  \label{fig:ipfs}
\end{figure*}

\begin{figure*}[htbp]
	\centering
	\includegraphics[width=0.95\textwidth]{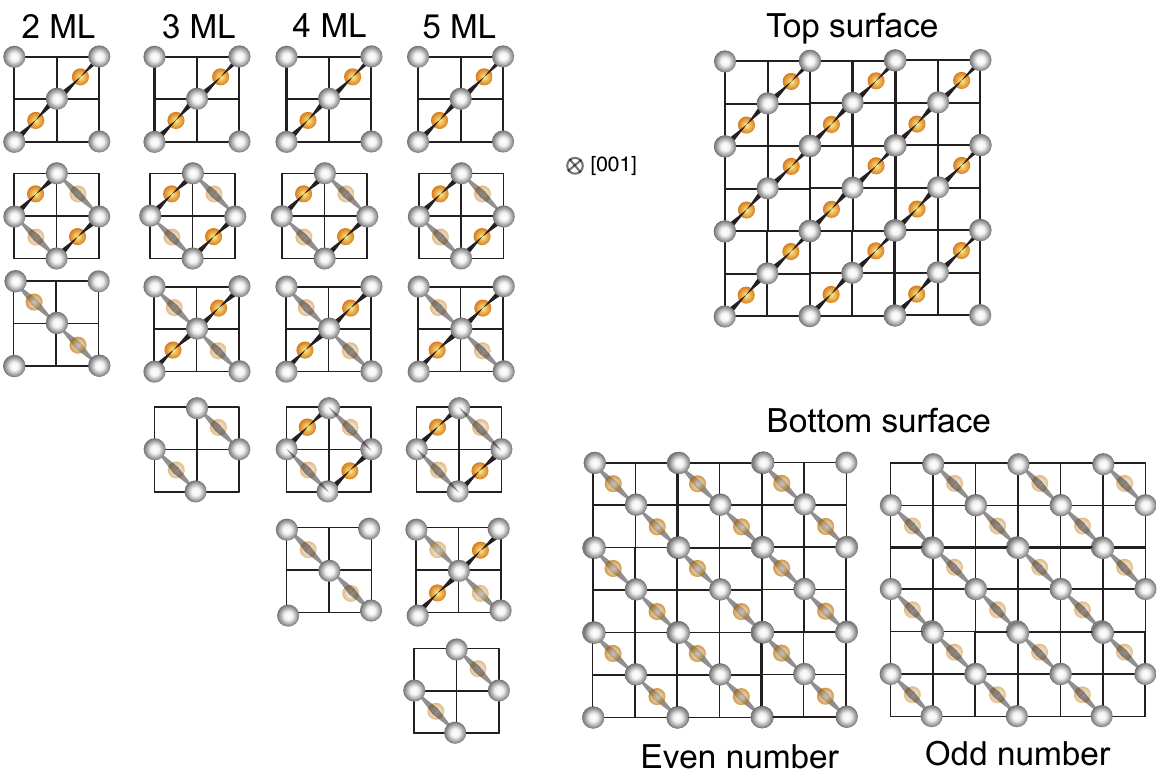}
  \caption{Crystallographic
      The structure of CdSe zinc-blende NPL depends on the number of monolayers viewed from the [001] direction. The atomic configurations for 2 to 5 monolayers are displayed on the left. On the right, the top and bottom surfaces show perpendicular Cd-Se bonding patterns for an even and an odd number of monolayers.}
  \label{fig:odd-even}
\end{figure*}

\begin{figure*}[htbp]
	\centering
	\includegraphics[width=0.95\textwidth]{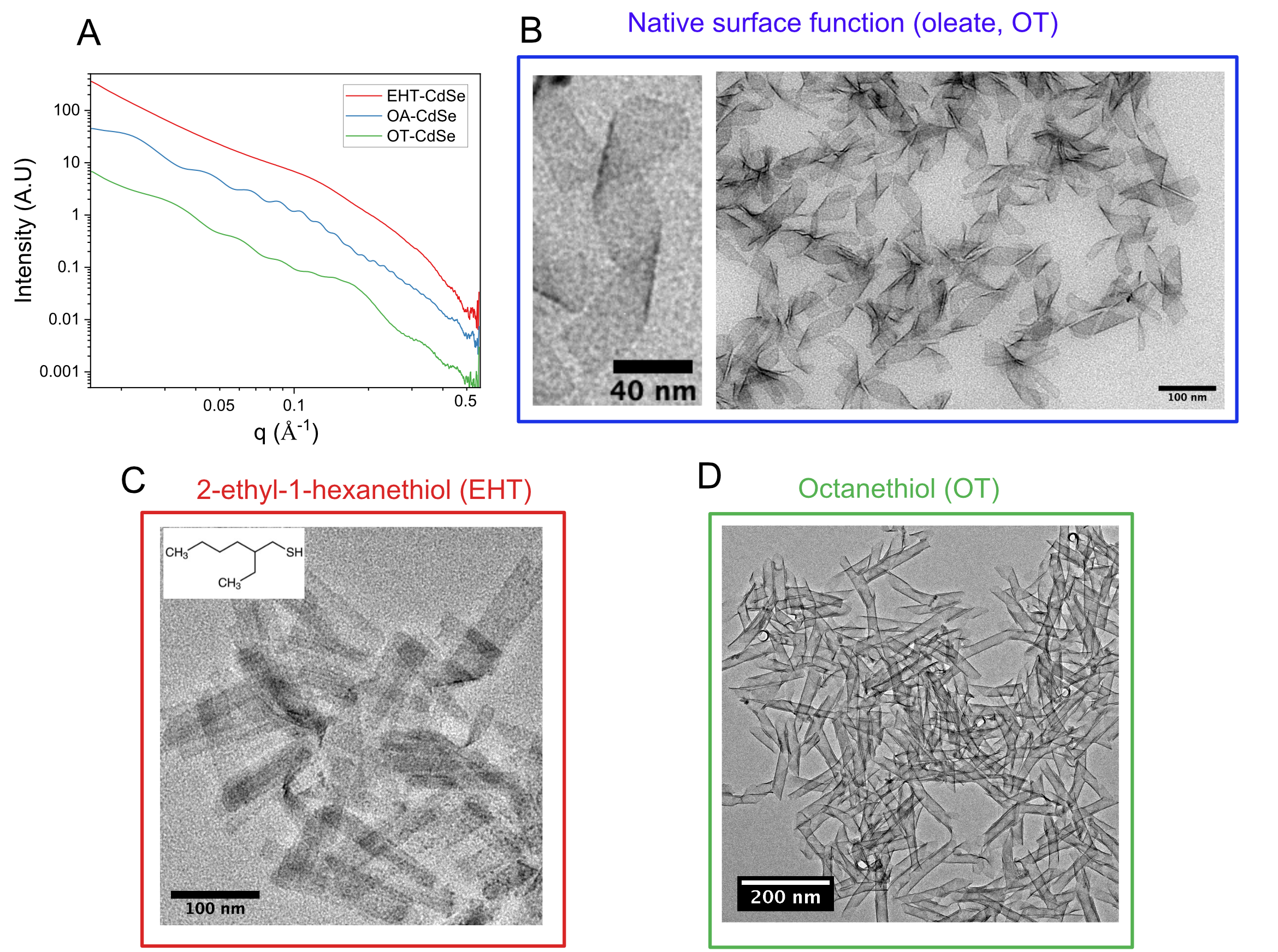}
    \caption{Effect of ligand chain structure on the curvature. A) SAXS patterns of CdSe NPL dispersion with different surface functions: oleate (native surface function), octanethiol and 2-ethyl-1-hexanethiol (EHT). The $q^{-2}$ slope indicates a flat NPL geometry. The oscillations in the oleate and octanethiol patterns indicate a curved surface. B) TEM images of CdSe NPL showing a helical ribbon geometry. C) TEM image of the same NPL after ligand exchange with (EHT) showing the unfolding of the helical ribbon into flat NPL D) TEM image of the same NPL after ligand exchange with octanethiol showing a curved geometry.}
	\label{fig:branching}
\end{figure*}

\end{document}